\documentclass[aps,prb,reprint,twocolumn,superscriptaddress,showpacs,floatfix]{revtex4-1}
\usepackage{hyperref}
\usepackage{graphicx}
\usepackage{amsmath}
\usepackage{amssymb}
\usepackage{color}
\usepackage{enumitem}
\usepackage{nicefrac}
\usepackage{multirow}

\newcommand{\beginsupplement}{%
        \setcounter{table}{0}
        \renewcommand{\thetable}{S\arabic{table}}%
        \setcounter{figure}{0}
        \renewcommand{\thefigure}{S\arabic{figure}}%
     }

\begin{document}

\title{Spin-unrestricted self-energy embedding theory}
\date{\today}
\author{Tran Nguyen Lan}
\email{latran@umich.edu}
\affiliation{Department of Chemistry, University of Michigan, Ann Arbor, Michigan, 48109, USA}
\altaffiliation{Department of Physics, University of Michigan, Ann Arbor, Michigan, 48109, USA}
\altaffiliation{Ho Chi Minh City Institute of Physics, VAST, Ho Chi Minh City, Vietnam}
\author{Sergei Iskakov}
\affiliation{Department of Physics, University of Michigan, Ann Arbor, Michigan, 48109, USA}
\author{Dominika Zgid}
\email{zgid@umich.edu}
\affiliation{Department of Chemistry, University of Michigan, Ann Arbor, Michigan, 48109, USA}

\begin{abstract}
\noindent We present a new theoretical approach, unrestricted self-energy embedding theory (USEET) that is a Green's function embedding theory  used to study problems in which an open, embedded system exchanges electrons with the environment. USEET has a high potential to be used in studies of strongly correlated systems with odd number of electrons and open shell systems such as transition metal complexes important in inorganic chemistry. In this paper, we show that USEET results agree very well with common quantum chemistry methods while avoiding typical bottlenecks present in these method.


\end{abstract}
\maketitle

In the theory community, the embedding theories such as dynamical mean field theory (DMFT)~\cite{Georges92,Georges96,savrasov_RevModPhys06}, self-energy embedding theory (SEET)~\cite{Zgid15,Tran15b,Tran16,Tran_GW_SEET,Tran_generalized_seet,zgid_njp17,simons_benchmark2}, self-energy functional theory (SFT)~\cite{Potthoff06,Potthoff03,sft_japan_2018}, or density matrix embedding theory (DMET)~\cite{dmet_knizia12,dmet_knizia13,DMET_bootstrap_jcp_2016,DMET_bootstrap_molphys_2017} are becoming increasingly popular since they offer a good compromise between cost and accuracy and highly accurate results can be delivered without exponential or even a high polynomial scaling, provided that the strongly correlated electrons can be somewhat decoupled from the weakly correlated electrons.
Additionally, these formalisms eliminate many problems present in typical theories for open shell systems such as complete active space self-consistent field (CASSCF) or complete active space perturbation theory second order (CASPT2)~\cite{roos1987complete,Kerstin:jpc/94/5483}, where two-, three and possibly  even four-body density matrices may be required for computations. In contrast, in the Green's function embedding approaches such as SEET or DMFT, only one-body Green's function is sufficient to carry out desired calculations. 

While these approaches are novel and certainly require some time be fully established in the quantum chemistry community, they have several appealing features. 
 First, all these embedding formalisms can be used without significant changes to treat both molecular and periodic problems making them appealing as a tool for studying periodic systems since they can deliver beside ground state electronic energy, also information about thermodynamics and spectra due to the inherent temperature dependence present in the Matsubara Green's function formalism. Moreover, these approaches are designed to keep the computational cost at bay since they naturally employ a separation of energy scales and only the strongly correlated electrons are treated by a high level method while the remaining, weakly correlated electrons are treated by a low level approach. 
If the number of strongly correlated electrons is too large and they cannot be all treated simultaneously, then schemes such a the generalized SEET approach~\cite{Tran_generalized_seet} can be used to solve such a problem employing a series of intersecting orbital subspaces. Finally, it is worth mentioning that the embedding approaches are a natural fit for extensions of current quantum chemistry methods to quantum computers and can be used in quantum-classical hybrid computation schemes. In these hybrids, the difficult and frequently exponentially  scaling part of the problem is solved by a quantum computer while the easier polynomial part is solved by a classical computer.

The Green's function embedding methods such as SEET are of particular interest to us since they offer a direct promise of treating periodic systems and materials science problems after a successful extension of this formalism to solid state is implemented. However, before such extensions can be implemented such a new method has to be validated against known quantum chemistry methods to develop best algorithmic schemes and ensure that all sources of errors are fully understood. Here, we focus on investigation of an implementation of SEET based on unrestricted reference that we will call unrestricted SEET (USEET).

Finite temperature Green's function methods are designed to deliver Green's functions that describe a grand canonical ensemble specified by a number of electrons. These methods cannot access a single excited state and are not spin-adapted since their aim is to describe an ensemble of excited states if necessary with different $S^2$ that are accessible at a given temperature. Accessing only the ground state is possible since at low temperatures for many systems the ground state is well energetically separated from the excited states. There are obvious advantages of such approaches in solids and for systems where the energetic separation between states is small, if one is interested in evaluating thermodynamic quantities  or simultaneously many states are accessible in an experiment. 
This stands in a stark contrast to traditional zero temperature methods that are usually spin-adapted and designed to illustrate a canonical ensemble and describe only a single state specified by number of electrons and multiplicity. 

Consequently, in Green's function methods an unrestricted implementation such as USEET allows us not only to treat systems with odd number of electrons but also access lowest excited states within a given $S_z$ sector, for example some of the high spin states. Moreover, the unrestricted implementation allows us to illustrate antiferromagnetic states since they are easily accessible when starting from an unrestricted solution.
This scheme is especially important since many chemically important systems such as transition metal complexes have an odd number of electrons. Moreover, many of the periodic systems display a spontaneous symmetry breaking and can be calculated by unrestricted methods since the $S^2$ symmetry in these systems is naturally not preserved. 
Consequently, in practice for any complicated large systems with an open shell character where an ensemble of states is accessible at a given finite temperature, an unrestricted Green's function method is a computational method of choice. In this paper, we carefully examine the first implementation of USEET and estimate the accuracy of this method.

USEET displays a number of significant differences when compared to  traditional SEET~\cite{Zgid15,Tran15b,Tran16,Tran_GW_SEET,Tran_generalized_seet,zgid_njp17,simons_benchmark2} based on a restricted reference (RSEET). 
Here, we list only the most important algorithmic steps while a detail description is presented in the Appendix~\ref{sec:appendix}. 
An USEET calculation can be started from unrestricted Hartree-Fock (UHF), unrestricted Green's function second order (UGF2)~\cite{Phillips15}, or unrestricted GW (UGW) that are performed on the entire systems. These methods provide 
spin-up and spin-down Green's functions, $\bf G^{\alpha \alpha}$ and $\bf G^{\beta \beta}$, as well as Fock matrices $\bf F^{\alpha \alpha}$ and $\bf F^{\beta \beta}$. If correlated methods such as GF2~\cite{Zgid14,Phillips15,Rusakov14,Rusakov16,Welden16,Kananenka15,Kananenka16,kananenka_hybrif_gf2,Iskakov_Chebychev_2018} or GW~\cite{Tran_GW_SEET} are used, we also obtain self-energies $\bf \Sigma^{\alpha \alpha}_\text{weak}$ and $\bf \Sigma^{\beta \beta}_\text{weak}$. 

Based on the information provided by a weakly correlated method for all orbitals present in the system we can build an average density matrix $\gamma=-[G^{\alpha \alpha}(\tau=1/(k_BT))+G^{\beta \beta}(\tau=1/(k_BT))]$, where $k_b$ is a Boltzmann constant and $T$ is a physical temperature. Subsequently, this matrix is diagonalized to yield average occupation numbers per orbital and averaged eigenvectors. We observed that similarly to RSEET these average occupation numbers provide a good starting estimation of weakly or strongly correlated character of an orbital and can be used to help with the active space selection. To the first approximation, a significant partial occupation indicate strongly correlated character of a given orbital. However, in some cases where a double shell has to be included in the active space to reach a high accuracy, orbitals that are almost doubly occupied or almost empty also need to be included within the active space.

Since the orbitals included in the active space can be numerous, in order to illustrate all interactions between these orbitals, we build multiple impurities $A_r$ based on the orbital grouping schemes used in generalized SEET and described in Ref.~\citenum{Tran_generalized_seet}. The basis transformation is performed for the impurity orbitals and all the impurity matrix elements $G^{\sigma \sigma}_{ij}, F^{\sigma \sigma}_{ij}, \Sigma^{\sigma \sigma}_{ij}$, where $\sigma=\alpha$ or $\beta$ and $i,j \in A_r$ are expressed in an orthogonal orbital basis. One particular choice here is the basis of natural orbitals (NOs). Similarly, we perform the transformation of the two-body integrals to the a priori chosen orthogonal basis. 
In this new orthogonal basis, we evaluate the spin-dependent hybridization function for the impurity orbitals as
\begin{equation}\label{eq:hyb}
\Delta^{\sigma \sigma}_{ij}=[(G^{\sigma \sigma})^{-1}]_{ij}-(i\omega_n+\mu^{\sigma})S^{\sigma \sigma}_{ij}+F^{\sigma \sigma}_{ij}+\Sigma^{\sigma \sigma}_{ij},
\end{equation}
where $S^{\sigma \sigma}$ is the overlap matrix, $\mu^\sigma$ is the spin-dependent chemical potential, and $\omega_n=(2n+1)k_BT$, is the Matsubara frequency grid discretization.
Subsequently, both the these spin-up and spin-down hybridizations are used to obtain spin-dependent bath parametrizations with $\epsilon^{\sigma}_p$, and $V^{\sigma \sigma}_{ip}$. The matrix elements describing one of the impurities $A_r$, $F^{\sigma \sigma}_{ij}, \epsilon^{\sigma}_p, V^{\sigma \sigma}_{ip}$ and 2-body integrals $v_{ijkl}$ where $i,j,k,l \in A_r$ are passed to the full configuration interaction (FCI)~\cite{iskakov2017exact} solver capable of treating spin-dependence of the one-body Hamiltonian. From this solver, we obtain $[G^{\sigma \sigma}_\text{imp}]_{ij}$ that with the help of the Dyson equation yields 
$[\Sigma^{\sigma \sigma}_\text{imp}]_{ij}$.
Subsequently, a new total spin dependent self-energies are set up as 
\begin{equation}\label{eq:sigma_tot}
[\Sigma^{\sigma \sigma}_\text{tot}]_{ij}=[\Sigma^{\sigma \sigma}_\text{imp}]_{ij}+[\Sigma^{\sigma \sigma}_\text{weak}]_{ij}-[\Sigma^{\sigma \sigma}_\text{dc}]_{ij},
\end{equation}
where $i,j\in A_r$. $\Sigma^{\sigma \sigma}_\text{dc}$ stands for a double counting correction that is subtracted explicitly and ensures that there is no double counting of electron correlation coming from the solver and a weakly correlated method.
For cases, where both, or at least one of the $i,j$ does not belong to the impurity, the $[\Sigma^{\sigma \sigma}_\text{tot}]_{ij}=[\Sigma^{\sigma \sigma}_\text{weak}]_{ij}$.
These new self-energies are used to construct an updated spin-dependent Green's function for the total system as
\begin{equation}\label{eq:G_tot}
G^{\sigma \sigma}_\text{tot}=[(i\omega_n+\mu^\sigma)S^{\sigma \sigma}-F^{\sigma \sigma}-\Sigma^{\sigma \sigma}_\text{tot}]^{-1}.
\end{equation}
For this Green's function the chemical potentials $\mu^\alpha$ and $\mu^\beta$ have to be found yielding a proper number of spin-up and soin-down electrons, respectively.
This new Green's function can be employed in a self-consistent manner to yield an improved hybridization. For details see Appendix. 

In the above described procedure, there are few details that are noteworthy. First, our algorithm contains an average density matrix that helps us to identify orbitals that should be included in the active space  and it provides a common basis of natural orbitals for both spin-up and spin-down components. This allows us to have a single orthogonal basis for 2-body integrals and these integrals do not need to be handled separately for the two spin components. While the memory and storage are doubled in the USEET procedure, the computational scaling remains unchanged since only a single impurity problem is solved.

To investigate the accuracy of the USEET algorithm as compared with standard wavefunction methods, we examined electronic structure of a series of examples such as a bond stretching in the OH radical and diatomic molecules containing transition metal atoms: CrH, MnH, FeO,  and CrO. The PySCF package\cite{sun2018pyscf} was used to generate all needed integrals as well as to perform standard quantum chemistry calculations such as HF, coupled cluster, CASSCF, and FCI.

\begin{figure} [!htb]\centering
  \includegraphics[width=\columnwidth]{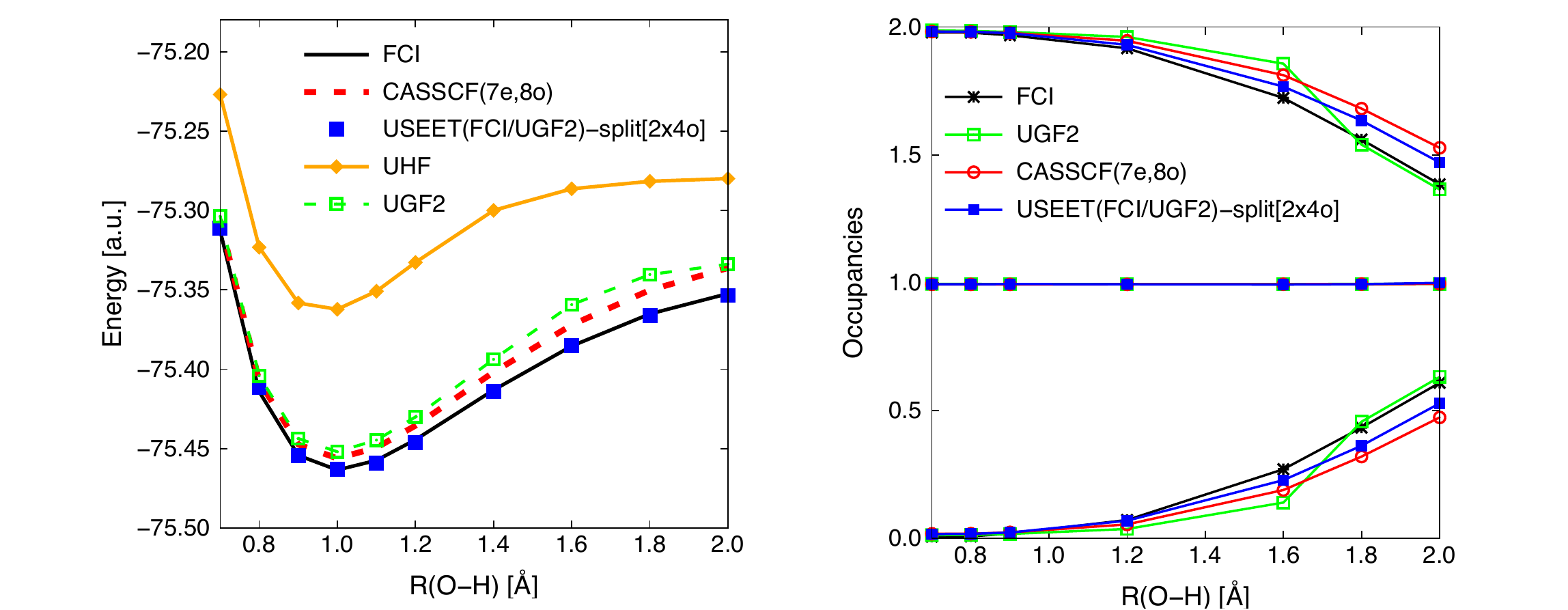}
  \caption{\normalsize Total energy (left panel) and occupation numbers (right panel) as functions of the OH bond length.}
  \label{fig:OH}
\end{figure}

We first consider a simple OH radical in the $^2\Pi$ ground state in the 6-31G basis \cite{Hehre:jcp/56/2257}. We have constructed a full active space of 8 NOs consisting majorly of O $2s2p3p$ and H $1s$ orbitals. The potential energy curves from different methods are summarized in the left panel of Figure~\ref{fig:OH}. As expected, UHF can describe the dissociation qualitatively, however, its curve is very far from the exact one. Although we have included 8 orbitals in active space from the total of only 11 orbitals, CASSCF(7e,8o) is still insufficient to recover FCI, indicating the importance of dynamic correlation. While UGF2 energies are quite good around the equilibrium  ( $ R < 1.2$ \r{A}), its nonparallelity error from FCI is significant (17 mH). 
For the USEET calculation, instead of using the full active space of 8 NOs, we split it into two groups according to the orbital symmetry: $\sigma$-type group (O $2s2p_z3p_z$ and H $1s$) and $\pi$-type group (O $2p_{x,y}3p_{x,y}$). Interestingly, USEET(FCI/UGF2)-split[2$\times$4o], in which both static and dynamic correlations are treated in a balanced way, excellently agrees with the FCI result.

To further explore the performance of USEET, in the right panel of Figure~\ref{fig:OH}, we plotted occupation numbers as a function of bond length for FCI, CASSCF, and USEET calculations. Although CASSCF occupation numbers follow FCI trend that smoothly shift from single-reference to multi-reference as the bond length increases, the deviation from the FCI occupation numbers are non-negligible at longer distances ( $ R > 1.2$ \r{A}). In UGF2, the inappropriate description of the molecular dissociation is reflected by a kink at $R = 1.6$ \r{A} resulting in a large nonparallelity error of the potential energy curve. When the static correlation is properly treated using the SEET(FCI/UGF2)-split[2$\times$4o] calculation, the transition becomes completely smooth and USEET occupation numbers are somewhat closer to FCI than CASSCF with the same number of active orbitals. 

\begin{figure} [!htb]\centering
  \includegraphics[width=\columnwidth,height=5.3cm]{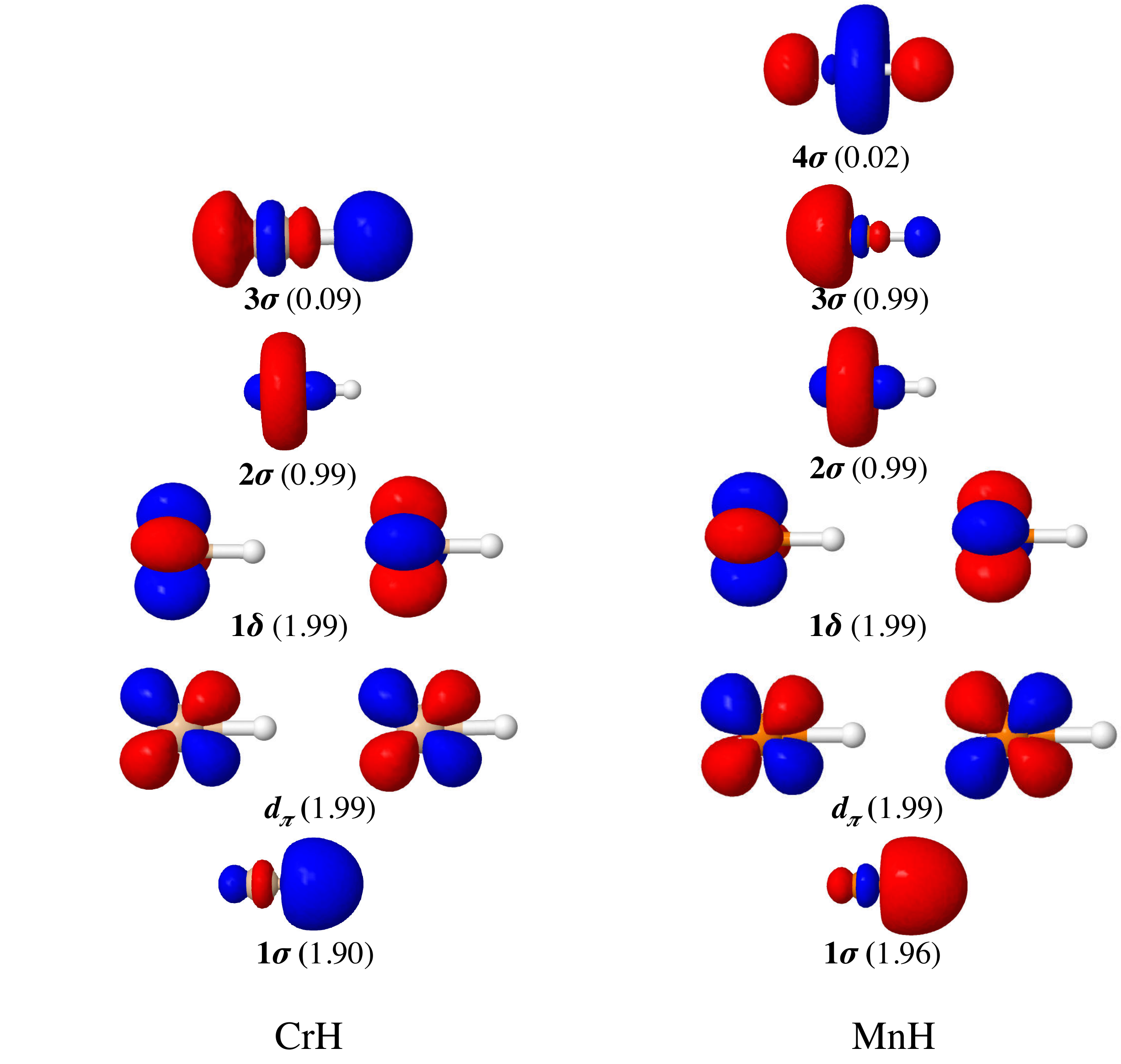}
  \caption{\normalsize Valence orbitals and their occupation numbers of CrH (left) and MnH (right) from USEET(FCI/UGF2) calculations.}
  \label{fig:hydrides}
\end{figure}

After the assessment of the simple OH radical, we consider simple transition metal hydrides: CrH and MnH. We have used experimental geometries \cite{bao2017predicting} for these hydrides: $R_{\mbox{\scriptsize{CrH}}} = 1.656$ \r{A} and $R_{\mbox{\scriptsize{MnH}}} = 1.731$ \r{A}. We have employed  energy consistent correlated electron pseudopotentials (eCEPPs) associated with aug-cc-pV$n$Z-eCEPP ($n$ = D and T) basis sets for Cr and Mn \cite{trail2017shape}. The cc-pV$n$Z ($n$ = D and T) basis sets were employed for H \cite{dunning_ccpvdz}. All orbitals used for USEET calculations are summarized in Figures~\ref{fig:S1} and \ref{fig:S2} in the supporting information (SI). 
Experiments and quantum chemical calculations \cite{harrison2000electronic} showed that the ground states of CrH and MnH are $^6\Sigma$ and $^7\Sigma$  states, respectively. While the ground state of CrH originates from $4s^1 3d^5$ of Cr and $1s^1$ of H giving rise to the $1\sigma^2 2\sigma^1 d_{\pi}^2 d_{\delta}^2 3\sigma^0$ configuration, the combination of $4s^2 3d^5$ of Mn and $1s^1$ of H leads to the $1\sigma^2 2\sigma^1 d_{\pi}^2 d_{\delta}^2 3\sigma^1 4\sigma^0$ configuration of MnH. According to occupation numbers displayed in Figure~\ref{fig:hydrides}, these ground state configurations are exactly reproduced by our USEET(FCI/UGF2) calculations.
Moreover, USEET yields CrH and MnH orbital character (see SI for the description) that are perfectly consistent with those obtained from previous multireference calculations \cite{dai1993spectroscopic,tomonari2009ab} showing that USEET is perfectly capable of achieving accuracy typical for common multireference methods.

\begin{table}[!h]
  \normalsize
  \caption{\label{tab:hydrides} \normalsize Binding energies (kcal/mol) of CrH and MnH. Experimental values were taken from Ref.~\citenum{bao2017predicting}. The cardinal number of the basis set is denoted as $n$.}
  \begin{tabular}{cccccccccc}
    \hline \hline		
    &$n$ &UHF	&CCSD	&UGF2	&USEET	&exp \\
    \hline	
\multirow{2}{*}{CrH}	&D	&24.9	&47.2	&46.2	&44.8	&\multirow{2}{*}{46.8} \\
	&T	&25.7	&49.6	&50.2	&47.5 \\
\multirow{2}{*}{MnH}	&D	&33.7	&36.0	&37.7	&29.9	&\multirow{2}{*}{31.1} \\
	&T	&33.9	&38.5	&37.3	&30.3 \\
    \hline \hline
  \end{tabular}
\end{table}

Let us now discuss binding energies of CrH and MnH. Despite relatively simple ground state electronic structure of CrH and MnH, their binding energies have been found to be challenging for density functional theory (DFT) and single-reference coupled cluster (CC) calculations \cite{xu2015practical,doblhoff2016diffusion,bao2017predicting}. In this work, the binding energy of MH (with M = Cr and Mn) was evaluated using the formula: $\Delta E =  E(\mbox{M}) + E(\mbox{H})- E(\mbox{MH})$, where $E(\mbox{M})$, $E(\mbox{H})$, and $E(\mbox{MH})$ are total energies of M and H atoms, and the MH molecule, respectively.
All calculated data are summarized in Table~\ref{tab:hydrides}.  
While UHF is fortuitously good for MnH, it badly fails for CrH. Interestingly, UGF2 results are comparable to the coupled cluster singles and doubles (CCSD) ones in all cases with the difference less than 2 kcal/mol. However, both these methods are still quite far from the experiment with the maximum error about 7 kcal/mol for MnH. A significant improvement upon UGF2 is achieved when USEET(FCI/UGF2)-11o and USEET(FCI/UGF2)-13o are performed for CrH and MnH, respectively. They result in an error  of less than 1 kcal/mol at the VTZ level.   

The next system we use to examine the USEET performance is the iron monoxide molecule, FeO. This system usually serves as a benchmark for understanding the chemical bonding between a transition metal and oxygen that play a key role in many chemical processes, such as biological transport and catalysis. The electronic structure of FeO is challenging due to a high density of low-lying states. In the past, numerous high-level theories and spectroscopic techniques have been employed to characterize its electronic structure. Those studies have revealed that valence shells of FeO are built from the combination of Fe $4s3d$ and O $2p$ shells giving rise to the quintet ground state associated with the $1\sigma^2 p_{\pi}^4 \delta^3 d_{\pi}^2 4s^1 2\sigma^0$ electronic configuration ($^5\Delta$).  The doubly occupied $1\sigma$ and empty $2\sigma$ orbitals are predominantly bonding and corresponding anti-bonding ones coming from Fe 3$d_{z^2}$ and O $2p_z$ atomic orbitals. 
While $p_{\pi}$ orbitals have bonding character and are  largely polarized towards O, $d_{\pi}$ orbitals display antibonding character and are polarized towards Fe. Orbitals $\delta$ and $4s$ are non-bonding orbitals of Fe. By transferring an electron from the $\delta$ orbital to the anti-bonding $2\sigma$ one, the septet excited state $^7\Sigma$ ($1\sigma^2 p_{\pi}^4 \delta^2 d_{\pi}^2 4s^1 2\sigma^1$) is reached. The most recent experiment has assigned the $^7\Sigma$ state energy to be 0.616 eV above the $^5\Delta$ state energy \cite{kim2015low}.   

\begin{figure} [!htb]\centering
  \includegraphics[width=\columnwidth,height=5.3cm]{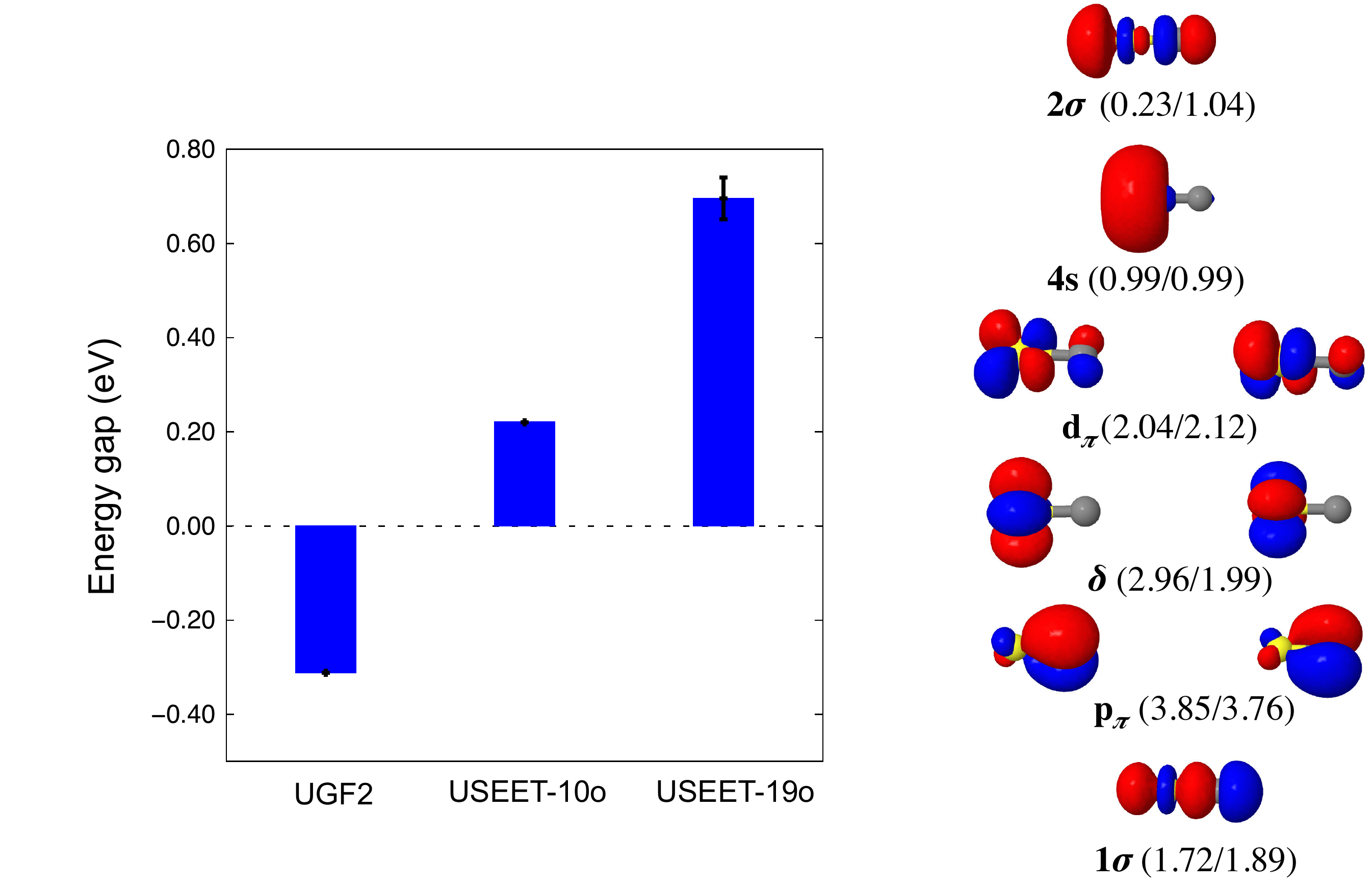}
  \caption{\normalsize Left panel: energy gap (eV) between septet and quintet states from UGF2 and USEET calculations. Right panel: valence orbitals and their occupation numbers for the quintet (first number) and septet (second number) states of FeO from the USEET-19o calculation. }
  \label{fig:FeO_gap}
\end{figure}

 We have employed eCEPPs associated with the aug-cc-pVDZ-eCEPP basis set for Fe and O \cite{trail2017shape}. The experimental FeO bond length of 1.616 \r{A}  \cite{bao2017predicting} was employed in our calculations. A single-shot UGF2 based on the UHF reference has been carried out instead of self-consistent UGF2 due to convergence problems. We have performed two USEET(FCI/UGF2) calculations. In the first one, only 10 valence orbitals were constructed from Fe $4s3d$ and O $2s2p$ atomic orbitals. It is well-known that double-shell effects are very significant for late transition metals, we therefore constructed a larger active space of 19 orbitals by further including Fe $4d$ and O $3s3p$ orbitals. We denote these two USEET(FCI/UGF2) calculations as USEET-10o and USEET-19o. All NOs used for USEET calculations are visualized in Figure~\ref{fig:S3} in SI.

The left panel of Figure~\ref{fig:FeO_gap} visualizes the energy gap between the quintet and the septet states of FeO. Note that in USEET, we evaluate quintet state as the state with $S_z = 4$ and septet state as the state with $S_z = 6$. There is a systematic error of 44 meV for USEET-19o corresponding to the convergence threshold of this calculation for the septet state. UGF2 predicts the energy of septet state --0.312 eV lower than the quintet state one. This result disagrees with experiments. The USEET(FCI/UGF2) calculation with 10 orbitals (USEET-10o) improves the UGF2 result and describes a correct ordering of these states, reflected by a positive energy gap (0.239 eV). It is, however, much smaller than the experimental value. When double-shell effects were taken into account in USEET-19o, the energy gap is enlarged to 0.696 eV, which is close to the experimental value (0.616 eV). The same trend was previously reported using the restricted active space second-order perturbation theory (RASPT2) \cite{vancoillie2011multiconfigurational}. 

An error of 80 meV is possibly introduced because we are using the same geometry, which was measured for the quintet ground state, to describe both states. To improve the energy gap, an optimized geometry of the septet state should be employed. Another source of error is most likely coming from the incompleteness of basis set. 
In order to get an insight into electronic configurations of quintet and septet states of FeO described by USEET-19o, we visualize the valence orbitals together with their occupation numbers in the right panel of Figure~\ref{fig:FeO_gap}. USEET-19o correctly illustrates $^5\Delta$ and $^7\Sigma$ electronic states of FeO. Moreover, orbital components are in an excellent agreement with RASPT2 calculations~\cite{vancoillie2011multiconfigurational}.

\begin{figure} [!htb]\centering
 \includegraphics[width=\columnwidth,height=5.3cm]{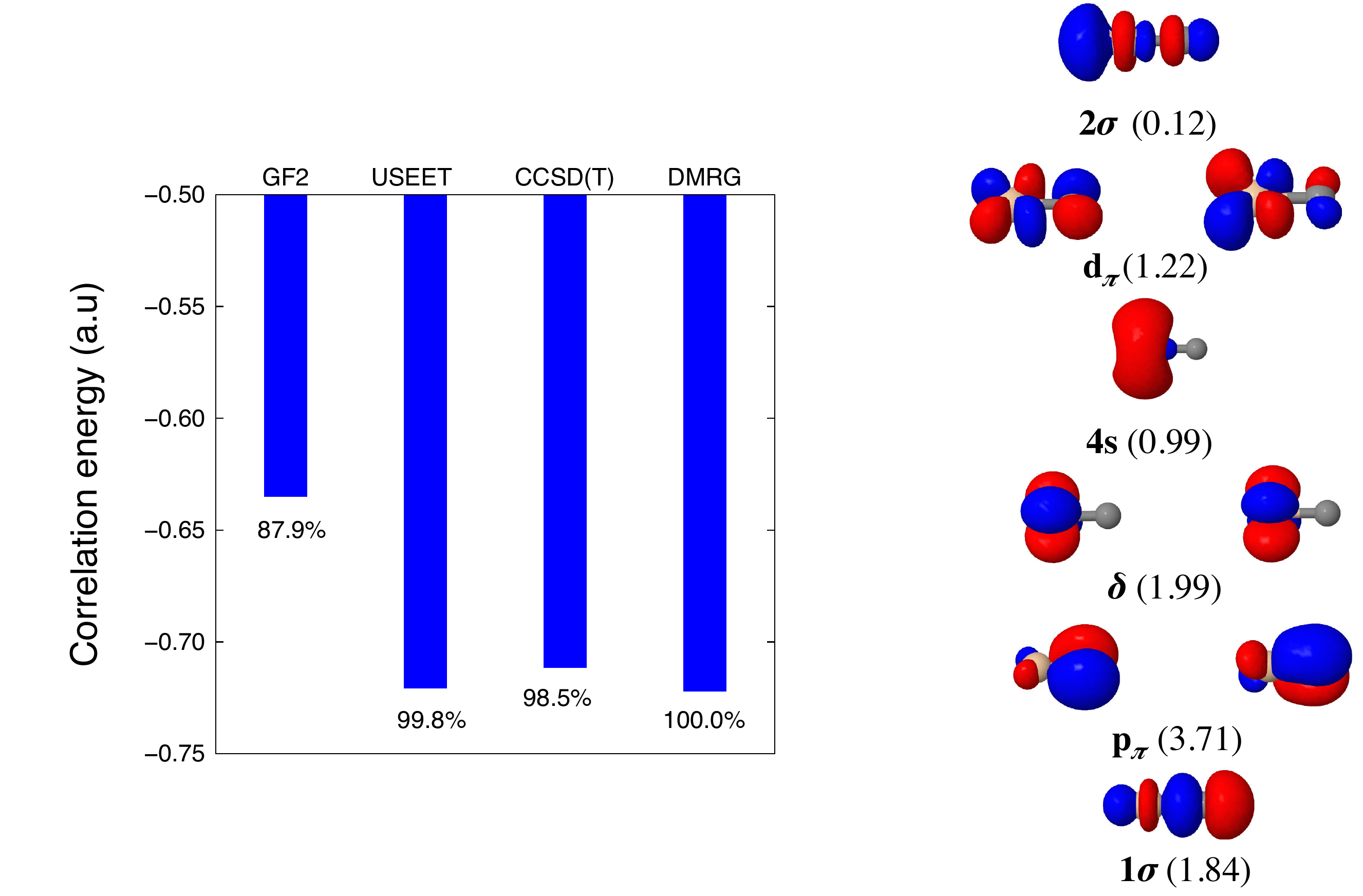}
  \caption{\normalsize Left panel: correlation energies from UGF2 and USEET calculations for the quintet state of CrO in comparison to CCSD(T) and DMRG ones. Right panel: valence orbitals and their occupation numbers for the quintet state of CrO from the USEET calculation.}
  \label{fig:CrO_corr}
\end{figure}

The last system we are going consider is a CrO molecule that is known to be very challenging due to its $^5\Pi$ ground state associated with the $1\sigma^2 p_{\pi}^4 \delta^2 d_{\pi}^1 4s^1 2\sigma^0$ electronic configuration resulting from a mixing of both covalent and ionic interactions \cite{jasien1988theoretical,harrison2000electronic}. The orbital components are basically the same as those of FeO.
The experimental bond length (1.621 \r{A}) was adapted from Ref.~\citenum{harrison2000electronic}.  We have employed eCEPPs associated with the aug-cc-pVDZ-eCEPP basis set for Cr and O \cite{trail2017shape}. The active space has been constructed from Cr $4s3d$ and O $2p$. The double-shell effect on Cr has been considered by further including three $\sigma$- and $\pi$-type Cr $4d$ orbitals giving rise to the full active space of 12 orbitals presented in Figure~\ref{fig:S4} in SI.

In Figure~\ref{fig:CrO_corr}, we are presenting correlation energies from UGF2, USEET(FCI/UGF2)-12o, coupled cluster singles and doubles with perturbative triples (CCSD(T)) calculations. We use the density matrix renormalization group (DMRG) \cite{White_DMRG_1992,Kurashige:jcp/130/234114,Reiher_ordering_2005,Chan_dmrg_2002,Sharma_Chan_review_2011,dmrg_block_2015} as a reference correlation energy (--0.7219 Ha),  
which is widely accepted to yield close to the exact correlation energy within a given basis. A single-shot UGF2 calculation can retrieve only 87.9\% of the DMRG correlation energy. CCSD(T) performs better recovering 98.5\% of the DMRG correlation energy with an error of 10.7 mH. When USEET is employed on top of UGF2, the DMRG correlation energy is recovered to an excellent degree with only a very small error of 1.4 mH (0.2\%). USEET also correctly describes the $^5\Pi$ ground state of CrO as can be seen from occupation numbers of valence orbitals displayed in the right panel of Figure~\ref{fig:CrO_corr}. An extensive study including USEET and benchmarking multiple methods for transition metal atoms and monoxides will be published in the near future by Simons Collaboration  on the Many-Electron Problem.~\cite{Simons_benchmark_2018}

We have presented the first implementation of SEET based on unrestricted reference and benchmarked our results against standard quantum chemistry methods. This implementation opens a new area for SEET calculations since it allows us to treat systems with an odd number of electrons,  access spin states with different $S_z$ values, and also provides a much more natural way of treating open shell problems and states that display antiferromagnetic character. 
This investigation of USEET is important since it allows us to gain additional information about Green's function methods that are a new class of embedding methods gaining popularity in quantum chemistry. 
While in traditional quantum chemistry, methods based on unrestricted reference often result in a large spin contamination, we speculate that in USEET such a contamination is minimized since the spin contamination appears in the UHF, UGF2, or UGW but then must be partially cured in the active space part of the calculation. Unfortunately, evaluating $S^2$ value in USEET is beyond scope of this paper since it requires an evaluation of a two particle Green's function and we will attempt it only in our future work.
We have demonstrated that USEET results compare well to highly accurate quantum chemistry methods such as CCSD(T), RASPT2, and DMRG but they do not require an evaluation of higher order density matrices thus removing memory bottlenecks present in typical multireference calculations. Moreover, an embedding technique such as USEET can be easily generalized to periodic problems, where a set of active space orbitals can be treated separately in every single cell.

D.Z and T.N.L acknowledge NSF grant CHE-1453894. S.I. was sponsored by the Simons Foundation via the Simons Collaboration on the Many-Electron Problem. T.N.L was also supported by Vietnamese National Foundation of Science and Technology Development (NAFOSTED) under grant 103.01-2015.14.  We thank Lucas Wagner for the Simons collaboration benchmark data. We are very grateful to Garnet Kin-Lic Chan and Sheng Guo for the DMRG energy for the CrO molecule.
\\
\appendix \label{sec:appendix}
\noindent {\bf Appendix}\\
Here, we list all the steps present in the USEET algorithm. The algorithm is composed from a lower level, external loop denoted here as {\bf LL1-LL6} steps and a higher level, internal loop denoted as 
{\bf HL1-HL7} steps. The higher level loop updates the coupling between spin unrestricted bath and the environment orbitals while the lower level loop is able to update the total unrestricted Green's function of the environment orbitals. The lower level loop from point {\bf LL6} can but does not need to be continued. 
\begin{description}
\item [LL1] Perform an unrestricted calculation, UHF, UGF2, or UGW and get 
$\bf G^{\sigma \sigma}, F^{\sigma \sigma}, \Sigma^{\sigma \sigma}_\text{weak}$, matrices, where $\sigma=\alpha$ or $\beta$. Initially, $\Sigma^{\sigma \sigma}_\text{tot}=\Sigma^{\sigma \sigma}_\text{weak}$
\item [LL2] \label{basis} Using the Green's function evaluate and an average density matrix $\gamma=-[G^{\alpha \alpha}(\tau=1/(k_BT))+G^{\beta \beta}(\tau=1/(k_BT))]$, where $k_B$ is the Boltzmann constant and $T$ is the physical temperature. The eigenvectors of this density matrix are creating a common basis of both spins that is orthogonal. Since the only requirement is that the basis for building the impurity problem and the embedding part of the algorithm is orthogonal, other bases such as basis of orthogonal localized orbitals, CASSCF orbitals, or even HF and DFT orbitals are also possible.
\item [LL3] With the help of the average density matrix choose the most strongly correlated orbitals (or orbitals that are most important to reach high accuracy) that will build an active space. Here, again using the average density matrix is only one choice, other criteria for selecting active space orbitals are also possible.
\item [LL4] Transform $\bf G^{\sigma \sigma}, F^{\sigma \sigma}, \Sigma^{\sigma \sigma}$ to the orthogonal basis determined in point~{\bf LL2}. Transform the subset of 2-body integrals belonging to the active space to the new orthogonal basis.
\item [LL5]  Split the active space into groups of impurity orbitals. This grouping can be executed according to a scheme discussed in Ref.~\citenum{Tran_generalized_seet}. This means that a large number of active space orbitals can be split into multiple smaller impurity problems.
\item [HL1]  For each of the impurity problems build a spin-up and spin-down hybridization function according to Eq.~\ref{eq:hyb}. For each of these hybridizations find a group of $\epsilon^{\sigma}_p$, and $V^{\sigma \sigma}_{ip}$ describing the impurity Hamiltonian. 
\item [HL2] Pass the impurity parametrization $\epsilon^{\sigma}_p$, and $V^{\sigma \sigma}_{ip}$, $F^{\sigma \sigma}_{ij}$, $\mu^{\sigma}$, and 2-body integrals $v_{ijkl}$ where $i,j,k,l \in A_r$ and $p$ counts bath orbitals, to the full configuration interaction (FCI)~\cite{iskakov2017exact} solver capable of treating spin-dependence of the one-body Hamiltonian. Do not forget to subtract the double counting term $[F^{\sigma \sigma}_\text{DC}]_{ij}=\sum_{kl \in A_r}\gamma^{\sigma \sigma}_{kl}(v_{ijkl}-v_{ilkj})$ from the Fock matrix.
\item [HL3] From the solver, obtain $[G^{\sigma \sigma}_\text{imp}]_{ij}$ that with the help of the Dyson equation yields $[\Sigma^{\sigma \sigma}_\text{imp}]_{ij}$.
\item [HL4] Set up a new total spin-dependent self-energy according to Eq.~\ref{eq:sigma_tot}
\item [HL5] Use this newly constructed self-energy to construct the Green's function for the total system according to Eq.~\ref{eq:G_tot}.
\item [HL6] Find the new chemical potentials $\mu^{\sigma}$ that yield a proper number of spin-up and spin-down electrons present in the system.
\item [HL7] Come back to point  {\bf HL1} and iterate until the hybridization will stop to change.
\item [LL6] If desired the low level loop can be continued here and using the new correlated Green's function from Eq.~\ref{eq:G_tot} a single iteration of UHF, UGF2, or GW can be performed here yielding new $\bf G^{\sigma \sigma}, F^{\sigma \sigma}, \Sigma^{\sigma \sigma}_\text{weak}$ matrices. The algorithm can be continued from point {\bf LL2} until the convergence of Green's function is achieved.

\end{description}


%

\clearpage
\onecolumngrid

\section*{Supporting Information}
\beginsupplement
\subsection*{Orbital components of CrH and MnH:} 

See the main text for the notation describing molecular orbitals. For CrH, while $1\sigma$ is a bonding orbital with main contribution from H $1s$, the $3\sigma$ is an anti-bonding orbital built from Cr $d_{z^2}$ and H $1s$ orbitals. Orbital $2\sigma$ is a predominantly non-bonding one coming from Cr $4s+3d_{z^2}$. Orbitals $d_{\pi}$ and $d_{\delta}$ are pure $3d$ orbitals of Cr. 
For MnH, components of $1\sigma$, $2\sigma$, $d_{\pi}$, and $d_{\delta}$ orbitals are basically the same as those of CrH. Orbital $3\sigma$ is an anti-bonding orbital built mainly from Mn $4s3d_{z^2}$  with a minor contribution form H $1s$. This orbital is  largely polarized toward Mn. The empty $4\sigma$ orbital is an anti-bonding one built from  Mn $4s4p_z$ and H 1s. 

\subsection*{Active orbitals used for USEET calculations:}

\begin{figure} [h]
  \includegraphics[width=12.0cm,height=8.0cm]{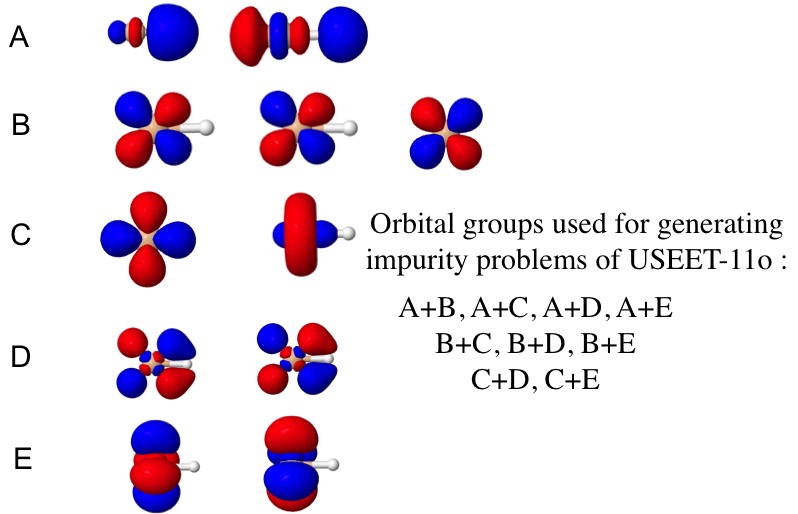}
  \caption{\normalsize Active orbitals  used for USEET-11o calculations of CrH.}
  \label{fig:S1}
\end{figure}

\begin{figure} [h]
  \includegraphics[width=12cm,height=9cm]{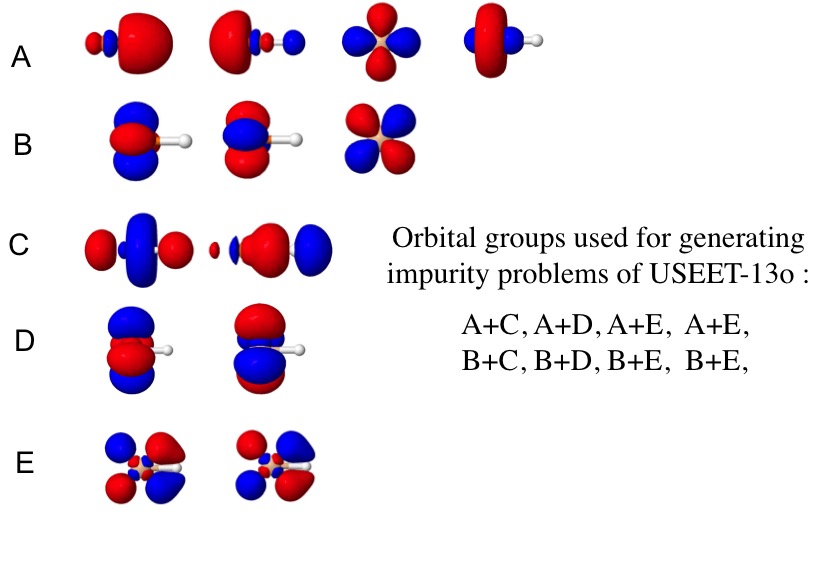}
  \caption{\normalsize Active orbitals used for the USEET-13o calculation of MnH.}
  \label{fig:S2}
\end{figure}

\begin{figure} [h]
  \includegraphics[width=12.0cm,height=13cm]{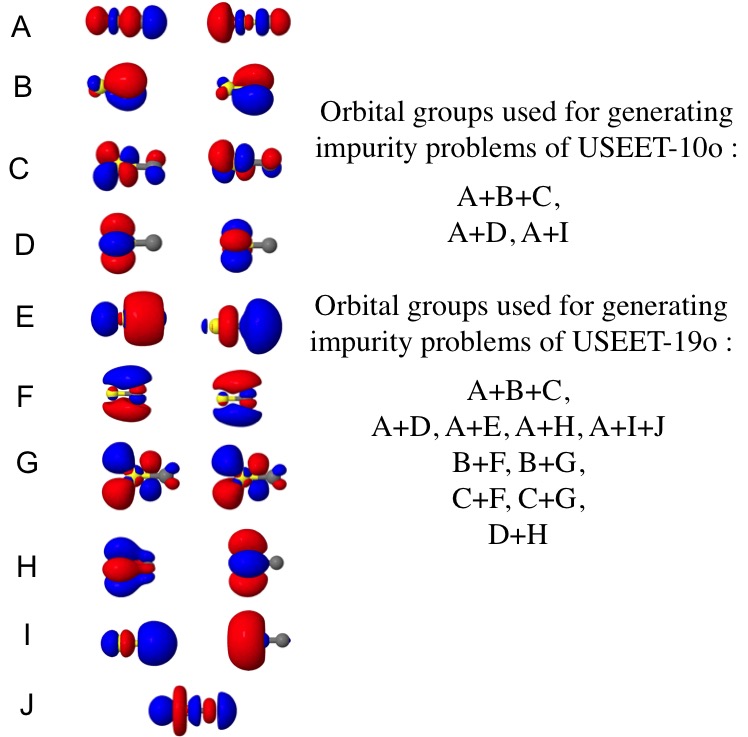}
  \caption{\normalsize Active orbitals used for USEET-19o calculations of FeO.}
  \label{fig:S3}
\end{figure}

\begin{figure} [h]
  \includegraphics[width=12cm,height=10cm]{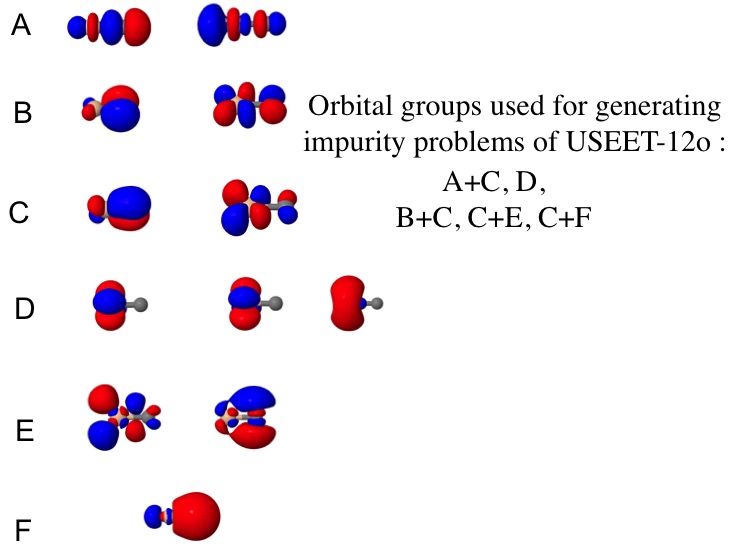}
  \caption{\normalsize Active orbitals used for USEET-12o calculations of CrO.}
  \label{fig:S4}
\end{figure}

\end{document}